\documentclass[a4paper,11pt]{article}

\usepackage[utf8]{inputenc}
\usepackage{amsfonts,amssymb,amsmath,amscd}
\usepackage{comment}
\usepackage{longtable}
\usepackage{color}
\usepackage{bm}
\usepackage{blkarray}
\usepackage{multirow}
\usepackage{accents}
\usepackage{graphicx}

\def\C{\mathbb{C}}
\def\R{\mathbb{R}}

\def\Z{\mathbb{Z}}
\def\M{\overline{M}}

\def\vxi{\vec{\xi}}
\def\veta{\vec{\eta}}
\def\prod{\,.\,}

\def\i{i}

\def\z{\mathrm{z}}

\def\dim{4}

\newcommand{\ROT}[2]{\Omega_{#1 , #2}}

\def\MTR{P}            

\def\MSC{K}            
\def\MDL{D}            

\newcommand{\MROT}[2]{J_{#1 , #2}}   

\def\ETR{T}           

\def\EDL{H}              
\def\ESC{C}               
\newcommand{\EROT}[2]{\Omega_{#1 , #2}}  

\newcommand{\E}[2]{E^{#1}_{#2}}

\def\ETH{\boldsymbol{\chi}}
\def\EBTH{\ovline{\ETH}}

\newcommand{\Eth}[2]{\chi^{#1}_{#2}}   

\newcommand{\Ebth}[2]{\ovline{\chi}_{#1}^{#2}}   

\newcommand{\EST}[2]{Q_{#1}^{#2}}   

\newcommand{\EBST}[2]{\ovline{Q}^{#1}_{#2}}   

\newcommand{\ESSC}[2]{S^{#1}_{#2}}   

\newcommand{\EBSS}[2]{\ovline{S}_{#1}^{#2}}   

\def\RSYM{\mathcal{A}}
\def\RCHA{\mathcal{R}}

\def\MTH{\boldsymbol{\theta}}
\def\MBTH{\ovline{\MTH}}

\newcommand{\Mth}[2]{\theta^{#1}_{#2}}   

\newcommand{\Mbth}[2]{\ovline{\theta}_{#1}^{#2}}   

\newcommand{\ETA}[2]{\eta_{#1 , #2}}

\newcommand{\DEL}[2]{\delta_{#1 , #2}}
\newcommand{\delt}[2]{\delta^{#1}_{#2}}

\newcommand{\com}[2]{\bigl[ #1 , \, #2 \bigr]}

\def \im{i}

\def\alp{\alpha}
\def\balp{\dot{\alpha}}
\def\bet{\beta}
\def\bbet{\dot{\bet}}
\def\bsigma{\dot{\sigma}}
\def\bgamma{\dot{\gamma}}

\def\Ia{A}
\def\Ib{B}

\def\Ic{C}

\newcommand{\ovline}[1]{\bar{#1}}

\def\esig{\text{\small $\Sigma$}}

\newcommand{\Esig}[3]{(\esig^{#1}){}_{#2}^{#3}}   

\newcommand{\EsigD}[3]{(\esig_{#1}){}_{#2}^{#3}} 

\newcommand{\EDsig}[3]{(\ovline{\esig}{}^{#1}){}^{#2}_{#3}}   

\newcommand{\EDsigD}[3]{(\ovline{\esig}{}_{#1}){}^{#2}_{#3}} 

\newcommand{\ERsig}[4]{(\esig_{#1,#2}){}^{#3}_{#4}}   

\newcommand{\ERsigU}[4]{(\esig^{#1,#2}){}^{#3}_{#4}}   

\newcommand{\ERDsig}[4]{(\ovline{\esig}_{#1,#2}){}^{#3}_{#4}}   

\newcommand{\ERDsigU}[4]{(\ovline{\esig}^{#1,#2}){}^{#3}_{#4}} 

\allowdisplaybreaks[1]

\def \TITLE {Compact Picture in Extended Superconformal Field Theories}

\newcommand{\Section}[1]{%
 \refstepcounter{section}
 \section*{\large \arabic{section}. #1}%
 \addtocontents{toc}{\protect\vspace{-8pt}}
 \addtocontents{toc}{\contentsline {section}{\thesection.\hspace{6pt}{#1}}{\arabic{page}}}}

\newcommand{\ASection}[1]{%
 \refstepcounter{section}
 \section*{\large Appendix \Alph{section}. #1}%
 \addtocontents{toc}{\protect\vspace{-8pt}}
 \addtocontents{toc}{\contentsline {section}{Appendix \thesection.\hspace{6pt}{#1}}{\arabic{page}}}}
\newcommand{\ASubsection}[1]{%
 \refstepcounter{subsection}
 \subsection*{\bf\normalsize \Alph{section}.\arabic{subsection}. #1}%
 \addtocontents{toc}{\protect\vspace{-10pt}}
 \addtocontents{toc}{\contentsline {section}{\rm\protect\hspace{16pt}\thesubsection.\hspace{3pt}{#1}}{\arabic{page}}}}

\def \setcntrs {\setcounter{equation}{0}}

\renewcommand{\thesection}{\arabic{section}}
\renewcommand{\theequation}{\arabic{section}.\arabic{equation}}

\title{Compact Picture in Extended\\ Superconformal Field Theories }

\author{Dimitar Nedanovski\thanks{dnedanovski@inrne.bas.bg}}

\date{}

\begin{document}

\maketitle

\thispagestyle{empty}

\vspace{-0.5cm}

{\footnotesize
\begin{center}
Institute for Nuclear Research and Nuclear Energy,\\
72 Tsarigradsko Chaussee Blvd., BG--1784 Sofia, Bulgaria
\end{center}
}
\begin{center}
\date{\today}
\end{center}

\vspace{0cm}

\begin{abstract}
There is a complex conformal transformation, which maps the $D$--di\-men\-sion\-al real Minkowski space on a bounded set in the 
$D$--di\-men\-sion\-al complex vector space. It generalizes the Cayley map from $D=1$ dimensions to higher space-time dimensions. This transformation provides a very convenient coordinate picture for Conformal Field Theories called compact picture. In this paper we extend the compact picture coordinates for superconformal field theories in four space-time dimensions.

\medskip

\noindent
Mathematics Subject Classification (2010): 181T99 , 81T60
\end{abstract}%

\tableofcontents

\Section{Introduction}
The Cayley map, $\R\ni t\mapsto \tfrac{t-i}{t+i}\in S^1$, which embeds the real line into the unit circle is convenient tool in the chiral (one-dimensional) conformal field theory. In analogy to this map, there is a complex conformal transformation \eqref{z->zeta} (for real $\zeta$) which maps Minkowski space $M\equiv\R^{3,1}$ into $(S^1\times S^3)/\Z_2=:\M$, that is isomorphic copy of the so called (conformally) compactified Minkowski space. The latter is a homogeneous space for the connected conformal group $SO_0(4,2)$. A quantum field theory with global conformal invariance (GCI QFT), as defined in \cite{NT01}, is naturally extended on $\M$. For the compactified complex Minkowski space there is a specific choice of coordinates which are global for the real $\M$. These provide a coordinate frame for the so called \textit{compact picture}, introduced and generalized in \cite{T86} and \cite{NT02}, respectively. In this work the term  ``compact picture'' will be used to refer to the above mentioned coordinates themselves.

The compact picture provides many useful features. First, there is an analog of the time coordinate - the conformal time. Its shift is a conformal transformation generated by the so called conformal Hamiltonian. As a consequence of the compactness, the latter has a discrete spectrum that corresponds to the scaling dimensions of the fields (since the conformal Hamiltonian is conjugated by a complex conformal transformation to the generator of dilatations in Minkowski space). 

There is another physical reason to consider a QFT in compact picture. It is related to the fact that  the universal covering of $\M$ is $\R\times S^3$ - the Einstein universe, which under certain limit process locally approximates the Minkowski space \cite{S71}.

After quantization of the theory one has to decompactify  the conformal time axis (the $S^1$ part of $\M$), but the spectrum of the the conformal Hamiltonian remains discrete and only its eigenvalues are changed, what corresponds to the so called anomalous scaling dimensions. This fact and the compactness of the space (the $S^3$ part of $\M$) facilitates the construction of perturbation theory. In particular, the infrared problems and the on-shell passage on a Hilbert space is expected to be
much more easy and manifest.

The compact picture has been used in \cite{N05} in the establishing of the equivalence between GCI QFT and the conformal vertex algebra in 4 dimensions, and latter in \cite{N15} where we propose a generalization for the case of superconformal symmetry.

To our knowledge, there is no perturbative QFT in compact picture. Probably this is related to the fact that for many field theories the conformal invariance which they have on a classical level is broken after quantization, and so we will not be able to pass to the real Minkowski space. However, for some superconformal field theories it has been proven that they remain perturbatively conformal invariant. (For example, $N=4$ super Yang-Mills theory \cite{SW81,M83}.) Thus, it should be possible their formulation in compact picture.


The paper is organized as follows. In the first section we make a short introduction of the compact picture, based on \cite{N05}.
The Second section describes the relation of complex Minkowski space and compact picture coordinates (identified with $\C^4$) 
to the Grassmannian $\mathrm{Gr}(2,\C^4)$. Detailed view of the identification of the Grassmannian coordinates  as a conformal compactification of the complex Minkowski space can be found in \cite[Sect.3.3]{V04}. In the third section we define a compact picture (CF. \eqref{flag_representative}) for the case with superconformal symmetry and give the transition (CF.  \eqref{transition_to_super_Minkowski}) to the complex super Minkowski space. 

We aim in future works to exploit the compact picture developed here  for a suitable formulation of the reality (unitarity)  conditions for superconformal field theories, and furthermore, provide a Hilbert space construction for these theories.
\setcounter{equation}{0}
\Section{Compact Picture on Minkowski Space}
The compact Minkowski space $\M$ is a homogeneous space for the conformal group $G\cong SO_0(4,2)$
\footnote{It is not a problem to set $G$ to be equal to the universal cover of $SO_0(4,2)$.}
in $\R^{3,1}$. By the Klein-Dirac construction it can be defined as $\M= Q/\R^*$, where
\begin{gather}\label{Compact_Minkowski_and_isotropic_cone}
 Q = \{ \vxi \in \R^{\dim,2}\backslash 0\ \vline \ \vxi^{\,2}\equiv \vxi.\vxi := - \xi_{-1}^2 - \xi_0^2 + \xi_1^2 + \xi_2^2 + \ldots + \xi_\dim^2=0 \}\,.
\end{gather}
$\M$ is characterized  by the stabilizer group $G_p$ of a point $p\in \M$.
The stabilizer is isomorphic to the Poincar\`{e} group with dilatations and have an Abelian normal subgroup,  $N_p$, isomorphic
to the group of translations.

Let, for $p\in\M$, $K_p$ be the \textit{light cone with tip at} $p$, i.e. the set of all rays in $\M$ orthogonal to the ray $p$, with the respect to the metric of $\R^{\dim,2}$. $K_p$ is invariant with respect to the action of $G_p$ and the action of $G_p$ splits $\M$ in two orbits: one closed, $K_p\backslash\{p\}$, and one open, $\M\backslash K_p$.  The relation $q \sim p$, defind as ``$q$ and $p$ are orthogonal'', is symmetric and $G$ -- invariant.

The group $N_p$ acts free and transitively on the open $G_p$ -- orbit $\M\backslash K_p$. Thus, we get an isomorphism 
\begin{equation}
 N_p\cong \M\backslash K_p, \quad t \mapsto t(q).
\end{equation}
Every pair of mutually nonisotropic (i.e. nonorthogonal)  points (rays) $q,p\in\M$ defines a chart in $\M$  \cite[Appendix A] {N05}. This chart is isomorphic to $N_p$. We will call $p$ ``a tip of the cone at infinity'' and $q$ - presents the center for the chart.

All these constructions are naturally transfered to the complex case, where $G$ is replaced by $G_{\C}\cong SO_0(6,\C)$ that is acting on the complexified compact Minkowski space $\M_{\C}=Q_{\C}/\C^*$, where $Q_{\C}$ is the complexification of \eqref{Compact_Minkowski_and_isotropic_cone}. In this case the stabilizer of a point $p\in \M_\C$ has an Abelian normal subgroup  $N_{\C,p} \cong \C^4$. In such a way one defines a special (analytic and algebraic) atlas on  $\M_{\C}$. The \textit{compact picture} is a chart of the above type, such that $p=q^*$, with respect to the conjugation $*$ that defines the real structure, and $q$ is chosen in the    \textit{future tube} \eqref{future_tube}.

Let $\{ \vec{e}_{-1}, \vec{e}_{0}, \vec{e}_{1}, \ldots, \vec{e}_{\dim} \}$ be the standard pseudo-orthonormal basis of $\C^{4, 2}$.
We shell define two charts, the first one will correspond to the complex Minkowski space, and the second one is the compact picture.

\medskip
\noindent \textbf{1.} Chart corresponding to complex Minkowski space:
\medskip

Let $p=\{\lambda \vxi_\infty\}$, where $\vxi_\infty:=\vec{e}_\dim - \vec{e}_{-1}$, be the ``tip of the cone at infinity'' and  $q=\{\lambda\vxi_0\}$, with
$\vxi_0:=\frac{1}{2} ( \vec{e}_\dim +  \vec{e}_{-1} )$, be the  ``center'' of  the chart. We construct the correspondence 

\begin{align}
M_{\C} \ni \zeta=\sum_{\nu = 0}^{3 }\zeta^\nu e_\nu \mapsto \vxi_{\zeta}  = 
                                       \frac{1 + \zeta^2}{2} \vec{e}_{-1} + \sum_{\mu = 0}^{3 } \zeta^{\mu} \vec{e}_{\mu} +\frac{1 - \zeta^2}{2} \vec{e}_{D},
                                       \label{zeta}
\end{align}
where $M_\C$ is the complex Minkowski space with $\{e_\mu\}_{\mu=0}^{3}$ a pseudo orthonormal basis in it. To a point $\{\lambda \vxi\} \in M_{p,q}$, with representative $\vxi$, corresponds  a point $\zeta\in M_\C$ with coordinates $\zeta^\mu= \frac{1}{\vxi_\infty . \vxi}\xi^\mu$.
($\vxi_\infty . \vxi\neq 0$, since by definition $\vxi\notin K_p$.)

\medskip
\noindent \textbf{2.} Compact picture coordinates:

\medskip
Let $v=\{\lambda \veta_\infty\}$, where $\veta_\infty:=\im \vec{e}_0 - \vec{e}_{-1}$, be the ``tip of the cone at infinity'' and  $w=\{\lambda\veta_0\}$, 
$\veta_0:=\frac{1}{2} ( \im \vec{e}_0 +  \vec{e}_{-1} )$, - the``center''.  It is clear that $ \veta_0$ and $\veta_\infty$ are complex conjugate to each other and  $\veta_0\sim \vxi_{\im e_0}$. The point $\im e_0\in M_\C$ is in the future tube\footnote{ The future tube \eqref{future_tube} and the past tube 
\begin{gather} 
 T^{-}:=\big\{ \zeta = x + \im y\,,\ \ x,y \in M\, \big| \, y^2<0\,\ \ -y^0 > 0 \big\}
\end{gather}
are homogeneous spaces for the connected real conformal group.
},
\begin{gather}\label{future_tube}
 T^{+}:=\big\{ \zeta = x + \im y\,,\ \ x,y \in M\, \big| \, y^2<0\,\ \ y^0 > 0 \big\}\,.
\end{gather}

The correspondence between this chart and $\C^4$ is
\begin{align}
\C^4 \ni \z=z^\mu f_\mu \mapsto \veta_{\z}  
                                 = \frac{1 + \z^2}{2} \vec{e}_{-1}  +\frac{1 - \z^2}{2} \i \vec{e}_{0} + \z^{\mu} \vec{e}_{\mu}\,,
                                 \label{compact_toC4}
\end{align}
where $\{f_\mu\}$ is an orthonormal $\C^4$ basis. Since in this paper we are mainly interested in the compact picture coordinates, we shell apply the Einstein summation rule for repeating upper and down indices running from one to four, like in \eqref{compact_toC4}. In the Minkowski space case with indices running form zero to three, we will explicitly write the summation, like in \eqref{zeta}. 

The transition between the charts 
is given by the  conformal transformation
\begin{gather}
 \Gamma=\exp( -\im \frac{\pi}{2} \ROT{0}{\dim})
 \label{Gamma_zeta_z_transform}
\end{gather}
followed by the change $ \zeta \mapsto (-\im z^D, z^1,\ldots,z^{\dim-1})$.
In terms of the coordinates $\zeta$ and $\z$, the transition functions are:
\begin{align}
\zeta^0  & = \frac{\i}{2} \frac{ 1 - \z^2}{ \frac{1 + \z^2}{2} + z^\dim}\,,              &         \boldsymbol{\zeta} & = \frac{\bm{\z}}{ \frac{1 + \z^2}{2} + z^\dim}\,,
\label{zeta->z}\\[1.25em]
z^\dim   & = \frac{1}{2} \frac{ 1 - \zeta^2}{ \frac{1 + \zeta^2}{2} - \i \zeta^0}\,,    &                    \bm{z}  & = \frac{\boldsymbol{\zeta}}{ \frac{1 + \zeta^2}{2} - \i \zeta^0}\,.
\label{z->zeta}
\end{align}

In the compact picture the real compactified Minkowski space, $\M$, corresponds to the set of points whose coordinates satisfy the reality condition $\z^*=\z$, where $\z^*:=\frac{\overline{\z}}{\overline{\z}^2}$ and $\overline{\z}=(\overline{z}^\mu)$ stands for componentwise complex conjugation of the vector $\z=(z^\mu)$. Note that 
\begin{gather}
 {\veta_{\z^*}}=\overline{(\tfrac{1}{\z^2}\,\veta_\z)}
 \label{conjugation_in_Klein_dirac}
\end{gather}
\setcounter{equation}{0}
\Section{Compact picture and Grasssmannian  model of $\M$}
The action of the conformal group $G$ on the Minkowski space $M$ is nonlinear. By the Klein-Dirac representation of $\M$  as $Q/R$ we linearize the problem and induce the nonlinear action from the linear action of $S0_0(4,2)$ on $\R^{4,2}$. This realization in the complex case further simplifies the conjugation of the compact picture coordinates. The  nonlinear formula is replaced in the Klein-Dirac realization  by the projective linear map \eqref{conjugation_in_Klein_dirac}.

In four-dimensional space-time there is another well known linearizing construction -- the Grassmannian model of $\M_\C$, which we will preview in this section with respect of the compact picture point of view. Here the realization of $\M_{\C}$ is given by the space of the complex plains in $\C^4$, i.e. the Grassmannian manifold $\mathrm{Gr}(2,\C^4)$. It is a homogeneous space for $SL(4,\C)$ and $\mathrm{Gr}(2,\C^4)\cong H_2\backslash SL(4,\C)$, where
\begin{gather}
H_2=\{ h\in SL(4,\C)\ | \ h \ \text{is of type} \left(\begin{smallmatrix}
                              A & 0_{2\times2}\\
                              B & C
                             \end{smallmatrix}\right)\}\,.
\end{gather}
Thus, the elements of $\mathrm{Gr}(2,\C^4)$ are right cosets
\begin{gather*}
 H_2\, \left(\begin{smallmatrix}
                              A_1 &  A_2\\
                              A_3 &  A_4
                             \end{smallmatrix}\right)\,,
\end{gather*}
$A_1,A_2,A_3,A_4$ are $2\times 2$--matrices, such that the block matrix that they form belongs to $SL(4,\C)$.
For $\z \in \C^4$ we introduce notation
\begin{gather}
  \widetilde{z}   :=  \sum_{k = 1}^{3} z^k \EDsig{k}{}{} + z^4 \EDsig{4}{}{}   = \begin{pmatrix}
                                                                             \im z^1 - z^2    & - \im z^3  +  z^4   \\
                                                                                  - \im z^3  -  z^4  & -  \im z^1 - z^2
                                                                                  \end{pmatrix}\,,
\end{gather}
with $\Sigma$ matrices given in \eqref{sigma_realization}.(The bar over the $\Sigma$-s does not mean complex conjugation.) $\C^4$ is identified with a chart in the Grassmannian:
\begin{align}\label{Compact_picture_Grassmannian}
 \C^4 \ni \z \mapsto H_2\, \begin{pmatrix}
                           \bm{1} & \widetilde{z}\\
                           \bm{0} & \bm{1}
                          \end{pmatrix}\,,
\end{align}
which corresponds to the compact picture from the previous section.

The  action of the proper conformal transformations in the compact picture now is given as a transposed right action of the corresponding elements of $SL(4,\C)$. Thus, to the transformation \eqref{z->zeta} corresponds the right multiplication of \eqref{Compact_picture_Grassmannian}  with the transposed of the matrix
\begin{gather}
\Gamma^{-1}
=
\exp(\im \frac{\pi}{2} \ROT{0}{\dim})
=
\frac{1}{\sqrt{2}}
\begin{pmatrix}
\bm{1}_{2\times 2} & -\alpha
\\
-\alpha &  \bm{1}_{2\times 2}
\end{pmatrix},\ \text{where}\quad
\alpha=\begin{pmatrix}
        0 & 1\\
        -1 & 0
       \end{pmatrix},
\nonumber
 \end{gather}
followed by the change $\z\mapsto(\zeta^1,\zeta^2,\zeta^3,\im \zeta^0)$\,. This gives transition to a chart
\begin{gather}
H_2\, \frac{1}{\sqrt{2}}\begin{pmatrix}
               \bm{1} + \widetilde{\zeta} \alpha &\alpha + \widetilde{\zeta}\\
               \alpha & \bm{1}
    \end{pmatrix}\,,\label{Grassmannian_Minkowski}
\end{gather}
identified with the complex Minkowski space. In \eqref{Grassmannian_Minkowski} we use the notion 
\begin{gather}
 \widetilde{\zeta}   := \im \zeta^0 \EDsig{4}{}{}  +\sum_{k = 1}^{3} \zeta^k \EDsig{k}{}{}     = \begin{pmatrix}
                                                                             \im \zeta^1 - \zeta^2    & - \im \zeta^3  +  \im \zeta^0   \\
                                                                                  - \im \zeta^3  -  \im\zeta^0  & -  \im \zeta^1 - \zeta^2
                                                                                  \end{pmatrix}\,.
\end{gather}                                                                                 

For $\det(\bm{1} + \widetilde{\zeta} \alpha )\equiv  \frac{1 + \zeta^2}{2} - \i \zeta^0 \neq 0$  we can write \eqref{Grassmannian_Minkowski} as
\begin{gather}
 H_2\, \begin{pmatrix}
  \bm{1} & \frac{\alpha + \widetilde{\zeta}}{\bm{1} + \widetilde{\zeta} \alpha }\\
               \bm{0} & \bm{1}
    \end{pmatrix}\,,
\end{gather}
thus  getting the relation
$\widetilde{z}=  \frac{\alpha + \widetilde{\zeta}}{\bm{1} + \widetilde{\zeta}\alpha }$  equivalent to \eqref{z->zeta}.
 
\setcounter{equation}{0}
\Section{Super Minkowski space and compact picture}
We aim to generalize the compact picture for the super Minkowski space in analogy to the previous section. We are using the setting of \cite{HH95}, where complexified compactified super Minkowski space is realized as a flag supermanifold, which is a homogeneous space for the complex superconformal group.

Consider the superspace $\C^{4|4N}$, where the even coordinates are the $\z$ coordinates of the compact picture and odd 
coordinates $\Eth{\alp}{A}$, $\Ebth{\balp}{A}$ (the bar used here is not a complex conjugation), $A=1,2,\ldots,N$, $\alp=1,2$, $\balp=\dot{1},\dot{2}$ are such that 
for a fixed $A$, they transform under $(1/2,0)$  and $(0,1/2)$ representations of $so(4)$, respectively. 

The identification \eqref{Compact_picture_Grassmannian} has a generalization in the case of extended superconformal symmetry.
Now, instead of the Grassmannian, we have flag supermanifold $F_{2|0,2|N}(4|N)$,  i.e. the space of all sequences of subspaces 
$\C^{2|0}$ $\subset$  $\C^{2|N}$ in $\C^{4|N}$. This flag supermanifold is a homogeneous superspace for the $SL(4|N,\C)$-the 
N-extended complex superconformal group:
$F_{2|0,2|N}(4|N)=H_{2|0,2|N}\backslash SL(4|N)$, where the stabilizer 
$H_{2|0,2|N}$ is $SL(4|N)$--subgroup of supermatrices with block structure 
\begin{gather}
 \left( \begin{tabular}{{cc|c}}
 $ A_{2\times 2}$ &$ 0_{2\times 2}$ & $0_{2\times N}$
\\
 $B_{2\times 2}$ & $C_{2\times 2}$ & $E_{2\times N}$
 \\
\hline
 $D_{N\times 2}$ & $0_{N\times 2}$ & $ F_{N\times N}$
\end{tabular} \right)\,.
\nonumber
\end{gather}
Let introduce the notations
\begin{gather}
\ETH:=\left(\begin{smallmatrix}
                  \Eth{1}{1} & \ldots & \Eth{1}{N}
                  \\
                  \Eth{2}{1} & \ldots & \Eth{2}{N}
                 \end{smallmatrix}\right)\,,
           \quad
           \EBTH:=\left(\begin{smallmatrix}
                   \Ebth{\dot{1}}{1} &  \Ebth{\dot{2}}{1}
                   \\
                   \vdots           &   \vdots
                   \\
                   \Ebth{\dot{1}} {N}&  \Ebth{\dot{2}}{N}
                  \end{smallmatrix}\right)\,,
          \quad
          \widetilde{Z}:=\widetilde{z} + 2 \ETH\EBTH\,.
          \nonumber
\end{gather}
Then $\C^{4\vline 4N}$ is identified with an open set
in $F_{2|0,2|N}(4|N)$ by the map
\begin{gather}
(z^1,z^2,z^3,z^4;\Eth{1}{1},\ldots, \Eth{1}{N},\Eth{2}{1},\ldots,\Eth{2}{N},\Ebth{\dot{1}}{1},\ldots,\Ebth{\dot{1}}{N},\Ebth{\dot{2}}{1},\ldots,\Ebth{\dot{2}}{N})
\nonumber\\
\rotatebox{-90}{$\longmapsto$}
\nonumber\\
H_{2|0,2|N}
 \left( \begin{tabular}{{c c |c}}
 $ \boldsymbol{1}_{2\times 2}$ &$ \widetilde{Z}$ & $-2\ETH$
\\
 $\boldsymbol{0}_{2\times 2}$ & $\boldsymbol{1}_{2\times 2}$ & $\boldsymbol{0}_{2\times N}$
 \\
\hline
 $\boldsymbol{0}_{N\times 2}$ & \raisebox{-0.3em}{$-2\EBTH$} & $ \boldsymbol{1}_{N\times N}$
\end{tabular} \right)\,.
\label{flag_representative}
\end{gather}
The corresponding action of $SL(4|N,\C)$ on $F_{2|0,2|N}(4|N)$ is given by right transposed matrix multiplication.
Using \eqref{compact_to_Omega}, \eqref{superconformal_to_sl(4)} and \eqref{matrix_realization_of_sl(4|N)_generators} we can 
write $ \Gamma^{-1}=\exp(\im \frac{\pi}{2} \ROT{0}{\dim})$ as a superconformal transformation:
\begin{gather}
\Gamma^{-1}= \left( \begin{tabular}{{cc|c}}
  \raisebox{0.2em}{${\scriptstyle{\frac{1}{\sqrt{2}}}}\boldsymbol{1}_{2\times 2}$} &
                                            \raisebox{0.2em}{$-{\scriptstyle{\frac{1}{\sqrt{2}}}}\alpha$} & {\multirow{2}{*}{$\boldsymbol{0}_{4\times N}$}} 
\\
$-{\scriptstyle{\frac{1}{\sqrt{2}}}}\alpha$ & ${\scriptstyle{\frac{1}{\sqrt{2}}}}\boldsymbol{1}_{2\times 2}$ &  
 \\
\hline
 \multicolumn{2}{c|}{$\boldsymbol{0}_{N\times 4}$} & $ \boldsymbol{1}_{N\times N}$
\end{tabular} \right)\ \in \ SL(4|N,\C)\,.
\end{gather}
Its action on \eqref{flag_representative}, followed by the change $\z\mapsto(\zeta^1,\zeta^2,\zeta^3,\im \zeta^0)$, 
$\Eth{\alp}{A}=\mathrm{e}^{\i\phi}\,\Mth{\alp}{A}$, $\Ebth{\balp}{A}=\mathrm{e}^{\i\phi}\,\Mbth{\balp}{A}$, $\phi\in \R$, gives
\begin{gather}
H_{2|0,2|N}
 \left( \begin{tabular}{{c c|c}}
 $ \frac{1}{\sqrt{2}}(\boldsymbol{1}_{2\times 2} +\widetilde{Z}_M\alpha)$  &   $ \frac{1}{\sqrt{2}}(\alpha + \widetilde{Z}_M)$ & $-2 \mathrm{e}^{\i\phi}\, \MTH$
\\
 $\frac{1}{\sqrt{2}}\alpha$ & $\frac{1}{\sqrt{2}}\boldsymbol{1}_{2\times 2}$ & $\boldsymbol{0}_{2\times N}$
 \\
\hline
 \raisebox{-0.3em}{$\frac{-2}{\sqrt{2}} \mathrm{e}^{\i\phi}\,\MBTH\alpha$}& \raisebox{-0.3em}{$\frac{-2}{\sqrt{2}}\mathrm{e}^{\i\phi}\, \MBTH$} & $ \boldsymbol{1}_{N\times N}$
\end{tabular} \right)\,.
\end{gather}
Thus, the transition to the extended super Minkowski space is 
\begin{gather}
 \widetilde{Z}
 =
 (1 +  \widetilde{Z}_M\alpha)^{-1} (\alpha + \widetilde{Z}_M)\,,
\quad
 \ETH 
 =
 -2\sqrt{2}(1 +  \widetilde{Z}_M\alpha)^{-1}\mathrm{e}^{\i\phi}\MTH\,,
\nonumber\\
\EBTH
 =
 \sqrt{2}
 \bigl(-4\mathrm{e}^{2\i\phi} \MBTH \alpha (1 +  \widetilde{Z}_M\alpha)^{-1} \MTH +1\bigr)^{-1}
 \mathrm{e}^{\i\phi}\MBTH
 \bigl( (1 +  \widetilde{Z}_M\alpha)^{-1}(\alpha + \widetilde{Z}_M) -1 \bigr)\,,
 \label{transition_to_super_Minkowski}
\end{gather}
where
\begin{gather}
\MTH:=\left(\begin{smallmatrix}
                  \Mth{1}{1} & \ldots & \Mth{1}{N}
                  \\
                  \Mth{2}{1} & \ldots & \Mth{2}{N}
                 \end{smallmatrix}\right)\,,
           \quad
           \MBTH:=\left(\begin{smallmatrix}
                   \Mbth{\dot{1}}{1} &  \Mbth{\dot{2}}{1}
                   \\
                   \vdots           &   \vdots
                   \\
                   \Mbth{\dot{1}} {N}&  \Mbth{\dot{2}}{N}
                  \end{smallmatrix}\right)\,,
          \quad
          \widetilde{Z}_M:=\widetilde{\zeta} + 2 \mathrm{e}^{2\i\phi}\MTH\MBTH\,.
          \nonumber
\end{gather}

\bigskip
\noindent {\bf Acknowledgements.}
The author thanks the Math Department at University of Geneva (where much of this work was complete) for the hospitality and great scientific environment. The directions and discussions (which go far behind this paper) given by Prof. Nikolay Nikolov are highly appreciated. This research has been supported by the Sciex-NMSch (Project 13.157) and Bulgarian NSF Grant
DFNI T02/6.

\setcounter{section}{0}
\renewcommand{\thesection}{\Alph{section}}
\renewcommand{\theequation}{\Alph{section}.\arabic{equation}}

\ASection{Notations and conventions}
\setcntrs
\ASubsection{Conformal algebra}
In our construction we have naturally four bases for the conformal Lie algebra: the $so(4,2)$ basis, two bases adapted to the conformal action in the chatrs described in Sect.2  and $sl(4)$ basis related to the conformal action on the Grassmannian in Sect.3.

\medskip

1. $so(4,2)$  basis for the conformal Lie algebra:
\begin{gather}
\com{\ROT{a}{b} } {\ROT{c}{d} }
=
\ETA{a}{c} \ROT{b}{d}
+
\ETA{b}{d} \ROT{a}{c}
-
\ETA{a}{d} \ROT{b}{c}
-
\ETA{b}{c} \ROT{a}{d}\,,
\\
\ROT{a}{b}=-\ROT{b}{a}\,,\  a,b,c,d=-1,0,1,\ldots,4\,, \ \eta=\mathrm{diag}(-1,-1,1,1,1,1)\,.
\nonumber\\
\intertext{For $a<b$, $\ROT{a}{b}$ are realized as matrices with elements }
(\ROT{a}{b})^c_d=\left\{\begin{array}{rcl}
                  -\delta_{ac}\delta_{bd} + \delta_{bc}\delta_{ad}, &\ \text{for}  &\ a,b=1,2,3,4\,;
                  \\
                  \delta_{ac}\delta_{bd} - \delta_{bc}\delta_{ad}, &\ \text{for} & \ a,b=-1,0\,;
                  \\
                  -\delta_{ac}\delta_{bd} - \delta_{bc}\delta_{ad}, &\ \text{for}  &\ a=-1,0,\ b=1,2,3,4\,.
                 \end{array}\right.
\nonumber               
\end{gather}

\medskip

2. For the complex Minkowski space:
\medskip

Generators $\MTR, \MSC, \MROT{}{}, \MDL$, which generate,respectively, the translations, special conformal transformations, rotations and dilatations in the complex Minkowski space.
\begin{gather}\label{ConCR_Minkowski}
\com{\MDL}{\MDL}
=
\com{\MDL}{\MROT{\mu}{\nu} }
=
\com{\MTR_\mu}{\MTR_\nu}
=
\com{\MSC_\mu}{\MSC_\nu}
=
0\,,
\nonumber\\[0.45em]
\com{\MROT{\mu}{\nu} } {\MROT{\rho}{\sigma} }
=
- \im \bigl(
\ETA{\mu}{\rho} \MROT{\nu}{\sigma}
+
\ETA{\nu}{\sigma} \MROT{\mu}{\rho}
-
\ETA{\mu}{\sigma} \MROT{\nu}{\rho}
-
\ETA{\nu}{\rho} \MROT{\mu}{\sigma}
\bigr)\,,
\nonumber\\[0.45em]
\com{\MTR_\mu}{\MSC_\nu}
=
-2 \im \bigl(
\ETA{\mu}{\nu} \MDL
-
\MROT{\mu}{\nu}
\bigr)\,,\nonumber
\end{gather}\vspace{-2.55em}
\begin{align}
\com{\MROT{\mu}{\nu}} {\MTR_{\gamma} }
=
- \im \bigl( 
\ETA{\mu}{\gamma} \MTR_{\nu}
-
\ETA{\nu}{\gamma} \MTR_{\mu}
\bigr)\,,
&\quad
\com{\MROT{\mu}{\nu}} {\MSC_{\gamma} }
=
- \im \bigl(
\ETA{\mu}{\gamma} \MSC_{\nu}
-
\ETA{\nu}{\gamma} \MSC_{\mu}
\bigr) \,,
\nonumber\\[0.45em]
\com{\MDL} {\MTR_{\mu}}
=
- \im
\MTR_{\mu}\,,
&\quad
\com{\MDL} {\MSC_{\mu}}
=
\im
\MSC_{\mu}\,,
\end{align}
$\mu,\nu,\rho,\gamma,\sigma=0,1,2,3\,.$

\medskip

3. For the compact picture chart:

\medskip

Generators $\ETR, \ESC, \EROT{}{}, \EDL$ -  generating, respectively, the translations, special conformal transformations, rotations and dilatations in the compact picture.
\begin{gather}\label{ConCR_Compact_Picture}
\com{\EDL}{\EDL}
=
\com{\EDL}{\EROT{\mu}{\nu} }
=
\com{\ETR_\mu}{\ETR_\nu}
=
\com{\ESC_\mu}{\ESC_\nu}
=
0\,,
\nonumber\\[0.45em]
\com{\EROT{\mu}{\nu} } {\EROT{\rho}{\sigma} }
=
\DEL{\mu}{\rho} \EROT{\nu}{\sigma}
+
\DEL{\nu}{\sigma} \EROT{\mu}{\rho}
-
\DEL{\mu}{\sigma} \EROT{\nu}{\rho}
-
\DEL{\nu}{\rho} \EROT{\mu}{\sigma}\,,
\nonumber\\[0.45em]
\com{\ETR_\mu}{\ESC_\nu}
=
2 \bigl(
\DEL{\mu}{\nu} \EDL
-
\EROT{\mu}{\nu}
\bigr)\,,\nonumber
\end{gather}\vspace{-2.55em}
\begin{align}
\com{\EROT{\mu}{\nu}} {\ETR_{\gamma} }
=
\DEL{\mu}{\gamma} \ETR_{\nu}
-
\DEL{\nu}{\gamma} \ETR_{\mu}\,,
&\quad
\com{\EROT{\mu}{\nu}} {\ESC_{\gamma} }
=
\DEL{\mu}{\gamma} \ESC_{\nu}
-
\DEL{\nu}{\gamma} \ESC_{\mu}\,,
\nonumber\\[0.45em]
\com{\EDL} {\ETR_{\mu}}
=
\ETR_{\mu}\,,
&\quad
\com{\EDL} {\ESC_{\mu}}
=
- \ESC_{\mu}\,,\label{conformal_CR}
\end{align}
$\mu,\nu,\rho,\sigma=1,2,3,4$.

\medskip

4. Realization as $sl(4,\C)$ algebra:
\begin{align*}
\com{\E{\alp}{\bet}}{\E{\gamma}{\sigma}}
&=
\delta^{\alp}_{\sigma}\,\E{\gamma}{\bet}\,-\,\delta^{\gamma}_{\bet}\,\E{\alp}{\sigma},
&\qquad
\com{\E{\balp}{\bbet}}{\E{\dot{\gamma}}{\dot{\sigma}}}
&=
\delta^{\balp}_{\dot{\sigma}}\,\E{\dot{\gamma}}{\bbet}
\,-\,
\delta^{\dot{\gamma}}_{\bbet}\,\E{\balp}{\dot{\sigma}},
\\[0.45em]
\com{\E{\alp}{\bet}}{\E{\dot{\gamma}}{\sigma}}
&=
\delta^{\alp}_{\sigma}\,\E{\dot{\gamma}}{\bet},
&\qquad
\com{\E{\balp}{\bbet}}{\E{\bgamma}{\sigma}}
&=
- \delt{\bgamma}{\bbet}\,\E{\balp}{\sigma},
\\[0.45em]
\com{\E{\alp}{\bet}}{\E{\gamma}{\bsigma}}
&=
-\,\,\delt{\gamma}{\bet}\,\E{\alp}{\bsigma},
&\qquad
\com{\E{\balp}{\bbet}}{\E{\gamma}{\bsigma}}
&=
\delt{\balp}{\bsigma}\,\E{\gamma}{\bbet},
\end{align*}\vspace{-1.55em}
\begin{gather}
\com{\E{\alp}{\bbet}}{\E{\bgamma}{\sigma}}
=
\delt{\alp}{\sigma}\,\E{\bgamma}{\bbet}
-
\delt{\bgamma}{\bbet}\,\E{\alp}{\sigma}\,, \label{sl_CR}
\intertext{and the generators satisfy}
\sum_{\alp} \E{\alp}{\alp}
+
\sum_{\balp} \E{\balp}{\balp}
=
0\,,\label{tr_sl_zero}
\end{gather}
$\alp,\bet,\gamma,\sigma=1,2,\ \balp,\bbet,\bgamma,\bsigma=\dot{1},\dot{2}$. The matrix realization of $\E{\alp}{\bet},\E{\balp}{\bbet},\E{\alp}{\bbet},\E{\balp}{\bet}$ is as in \eqref{matrix_realization_of_sl(4|N)_generators}, with the only difference that now the elementary matrices $e^i_j$ are $4\times 4\,.$

The relations of the generators used in point 1 to those used in points 2,3,4 are:
\begin{align}
 \im \MTR_\mu
&=
 - \ROT{-1}{\mu}
 -
   \ROT{\mu}{4}\,,
 &\quad
 \im \MDL
 &=
 \ROT{-1}{4}\,,&
\nonumber\\[0.45em]
 \im \MSC_\mu
 &=
 - \ROT{-1}{\mu}
 +
 \ROT{\mu}{4}\,,
 &\quad
 \im \MROT{\mu}{\nu}
 &=
 \ROT{\mu}{\nu}\,,\quad \mu,\nu=0,1,2,3.
\\[1.5em]
  \ETR_\mu
&=
 \im \ROT{0}{\mu}
 -
   \ROT{-1}{\mu}\,,
&\quad
\EDL
&=
 \im \ROT{-1}{0}\,,
\nonumber\\[0.45em]
 \ESC_\mu
&=
 - \im  \ROT{0}{\mu}
 -
 \ROT{-1}{\mu}\,,
&\quad
 \EROT{\mu}{\nu}
&=
 \ROT{\mu}{\nu}\,,\quad \mu,\nu=1,2,3,4.
\label{compact_to_Omega}
\end{align}
\begin{align}
 \E{\alp}{\bet}
 &=
 \frac{1}{2} \bigl( \im \delt{\alp}{\bet} \ROT{-1}{0} 
 -
 \ERsigU{\mu}{\nu}{\alp}{\bet} \ROT{\mu}{\nu} \bigr)\,,
 &\
 \E{\alp}{\bbet} 
 &=
 -  \frac{1}{2} \EDsig{\mu}{\alp}{\bbet} \bigl(  - \im  \ROT{0}{\mu} - \ROT{-1}{\mu} \bigr)\,,
\nonumber\\[0.45em]
\E{\balp}{\bbet}
 &=
 - \frac{1}{2} \bigl( \im \delt{\balp}{\bbet} \ROT{-1}{0} 
 -
\ERDsigU{\mu}{\nu}{\balp}{\bbet} \ROT{\mu}{\nu} \bigr)\,,
 &\
 \E{\balp}{\bet} 
 &=
 - \frac{1}{2} \Esig{\mu}{\bet}{\balp}  \bigl(  \im \ROT{0}{\mu} - \ROT{-1}{\mu} \bigr)\,,
 \label{sl4_to_conformal}
\end{align}
with $\Sigma$-matrices as given in \eqref{sigma_realization} and $\mu,\nu=1,2,3,4\,.$

An element of the connected with the unit component of the complex conformal group is given by 
$
\exp(\frac{1}{2}\sigma^{a,b}\ROT{b}{a})
$,
$
\exp\bigl(\im (\sum\limits_{\mu=0}^{3}p^\mu\MTR_\mu +\sum\limits_{\mu=0}^{3}k^\mu\MSC_\mu + \gamma \MDL +\frac{1}{2}\sum\limits_{\mu,\nu=0}^{3}j^{\mu,\nu}\MROT{\nu}{\mu})\bigr)
$,
$
\exp(t^\mu\ETR_\mu +c^\mu\ESC_\mu + h\EDL +\frac{1}{2}\omega^{\mu,\nu}\EROT{\nu}{\mu})
$ and
$
\exp(e^\alp_\bet \E{\bet}{\alp} +e^\alp_{\bbet} \E{\bbet}{\alp} + e^{\balp}_{\bet} \E{\bet}{\balp} + e^{\balp}_{\bbet} \E{\bbet}{\balp} )
$, corresponding to the realizations 1,2,3 and 4, respectfully.

\ASubsection{Extended superconformal Lie algebra}

a) Realization with set of generators adapted to the compact picture:

In addition to generators used in point 3. and their commutation relations \eqref{conformal_CR}, there are 
odd generators: $\EST{\alp}{\Ia}$, $\EBST{\balp}{\Ia}$, $\ESSC{\alp}{\Ia}$, $\EBSS{\balp}{\Ia}$;
even generators: $U(1)$-generator $\RCHA$  and $\RSYM^{\Ia}_{\Ib}$, $A,B=1,2,\ldots N$, spanning the Lie algebra  $sl(N,\C)$. 

The remaining nonzero commutation relation are:
\begin{align*}
 \com{ \EST{\alp}{\Ia}} { \EBST{\bbet}{\Ib}}
&=
2 
\, \delta^{\Ia}_{\Ib} \, \Esig{\mu}{\alp}{\bbet} \, \ETR_\mu \,,
& 
\com{ \ESSC{\alp}{\Ia}} { \EBSS{\bbet}{\Ib}}
&=
2 
\, \delta^{\Ib}_{\Ia} \, \EDsig{\mu}{\alp}{\bbet} \, \ESC_\mu \,,
&\\[1em]
\com{ \EROT{\mu}{\nu}  } { \EST{\alp}{\Ia} }
&=
\ERsig{\mu}{\nu}{\bet}{\alp} \ \EST{\bet}{\Ia}
\,, & 
\com{ \EROT{\mu}{\nu}  } { \EBST{\balp}{\Ia} }
&=
-
\ERDsig{\mu}{\nu}{\balp}{\bbet} \ \EBST{\bbet}{\Ia}
\,,
&\\
\com{ \EDL  } { \EST{\alp}{\Ia} }
&=
\frac{1}{2} \, \EST{\alp}{\Ia}
\,, & 
\com{ \EDL  } { \EBST{\balp}{\Ia} }
&=
\frac{1}{2} \, \EBST{\balp}{\Ia}
\,,
&\\
\com{ \ESC_{\mu}  } { \EST{\alp}{\Ia} }
&=
- 
\EsigD{\mu}{\alp}{\bbet} \, \EBSS{\bbet}{\Ia}
\,, & 
\com{ \ESC_{\mu}  } { \EBST{\balp}{\Ia} }
&=
\EsigD{\mu}{\bet}{\balp} \, \ESSC{\bet}{\Ia}
\,,
&\\
\com{ \RCHA  } { \EST{\alp}{\Ia} }
&=
\frac{1}{2} \, \EST{\alp}{\Ia}
\,,& 
\com{ \RCHA  } { \EBST{\balp}{\Ia} }
&=
-\frac{1}{2} \, \EBST{\balp}{\Ia}
\,,
&\\
\com{ \RSYM_{\Ic}^{\Ib}  } { \EST{\alp}{\Ia} }
&=
-(\delta^{\Ia}_{\Ic} \, \EST{\alp}{\Ib}-\frac{1}{N}\delta^{\Ib}_{\Ic}\,\EST{\alp}{\Ia})
\,,&
\com{ \RSYM_{\Ic}^{\Ib}  } { \EBST{\balp}{\Ia} }
&=
(\delta^{\Ib}_{\Ia} \, \EBST{\balp}{\Ic}-\frac{1}{N}\delta^{\Ib}_{\Ic}\,\EBST{\balp}{\Ia})
\,,
&\\[1em]
\com{ \ETR_{\mu}  } { \ESSC{\alp}{\Ia} }
&=
\EDsigD{\mu}{\alp}{\bbet} \, \EBST{\bbet}{\Ia}
\,, & 
\com{ \ETR_{\mu}  } { \EBSS{\balp}{\Ia} }
&=
- 
\EDsigD{\mu}{\bet}{\balp} \, \EST{\bet}{\Ia}
\,,
&\\
\com{ \EROT{\mu}{\nu}  } { \ESSC{\alp}{\Ia} }
&=
-
\ERsig{\mu}{\nu}{\alp}{\bet} \ \ESSC{\bet}{\Ia}
\,, & 
\com{ \EROT{\mu}{\nu}  } { \EBSS{\balp}{\Ia} }
&=
\ERDsig{\mu}{\nu}{\bbet}{\balp} \ \EBSS{\bbet}{\Ia}
\,,
&\\
\com{ \EDL  } { \ESSC{\alp}{\Ia} }
&=
-\frac{1}{2} \, \ESSC{\alp}{\Ia}
\,, & 
\com{ \EDL  } { \EBSS{\balp}{\Ia} }
&=
-\frac{1}{2} \, \EBSS{\balp}{\Ia}
\,,
&\\
\com{ \RCHA  } { \ESSC{\alp}{\Ia} }
&=
- \frac{1}{2} \, \ESSC{\alp}{\Ia}
\,, & 
\com{ \RCHA  } { \EBSS{\balp}{\Ia} }
&=
\frac{1}{2} \, \EBSS{\balp}{\Ia}
\,,
&\\
\com{ \RSYM_{\Ic}^{\Ib}  } { \ESSC{\alp}{\Ia} }
&=
(\delta^{\Ib}_{\Ia} \, \ESSC{\alp}{\Ic}-\frac{1}{N}\delta^{\Ib}_{\Ic}\, \ESSC{\alp}{\Ia})
\,,&
\com{ \RSYM_{\Ic}^{\Ib}  } { \EBSS{\balp}{\Ia} }
&=
-(\delta_{\Ic}^{\Ia} \, \EBSS{\balp}{\Ib}-\frac{1}{N}\delta^{\Ib}_{\Ic}\, \EBSS{\balp}{\Ia})
\,,&
\end{align*}
\begin{gather*}
\com{ \RSYM^{\Ia}_{\Ib}  } { \RSYM^{\Ia_1}_{\Ib_1} }
\, = \,
\delta^{\Ia}_{\Ib_1} \, \RSYM^{\Ia_1}_{\Ib}
\, - \,
\delta^{\Ia_1}_{\Ib} \, \RSYM^{\Ia}_{\Ib_1}\,,
\
\mathop{\sum}\limits_{\Ia=1}^N\RSYM^{\Ia}_{\Ia}=0\,,
\end{gather*}
\begin{align}
\com{ \EST{\alp}{\Ia}  } { \ESSC{\bet}{\Ib} }\,
&= \,
-2\,
\delta^{\Ia}_{\Ib}\,
\bigl(\delta_{\alp}^{\bet}\,\EDL
-
\ERsigU{\mu}{\nu}{\bet}{\alp} \, \EROT{\mu}{\nu}\bigr)
-4\,\delta_{\alp}^{\bet}
\RSYM_{\Ib}^{\Ia}
+ 2 \Bigl( \frac{4}{N}-1\Bigr) 
\delta^{\Ia}_{\Ib}\,\delta_{\alp}^{\bet}
\, \RCHA\,,
\nonumber\\
\com{ \EBST{\balp}{\Ia}  } { \EBSS{\bbet}{\Ib} }\,
&= \,
2\,\delta_{\Ia}^{\Ib}\,
\bigl(\delta_{\bbet}^{\balp} \, \EDL
+
\ERDsigU{\mu}{\nu}{\balp}{\bbet} \, \EROT{\mu}{\nu}{\big)}
-4\,\delta_{\bbet}^{\balp}
\RSYM_{\Ia}^{\Ib}
+ 2 \Bigl( \frac{4}{N}-1\Bigr)  
\delta_{\Ia}^{\Ib}\,\delta_{\bbet}^{\balp}
\, \RCHA\,,
\end{align}
with $\Sigma$-matrices defined as:
\begin{align*}
\Esig{\mu}{\bet}{\balp} 
&=
(\bar{\sigma}^{\mu}_{E})^{\balp\hspace{1pt}\gamma}\varepsilon_{\gamma\hspace{1pt}\bet}\hspace{1pt}\,,
&\
\ERsig{\mu}{\nu}{\gamma}{\alpha}
&=
-\,\frac{1}{4}\bigl(\,\EDsigD{\mu}{\gamma}{\bbet}\EsigD{\nu}{\alp}{\bbet}        
- \EDsigD{\nu}{\gamma}{\bbet}\EsigD{\mu}{\alp}{\bbet}\,\bigr)\,,
&\\
\EDsig{\mu}{\alp}{\bbet}
&=
\varepsilon^{\alp\hspace{1pt}\gamma}(\sigma^{\mu}_{E})_{\gamma\hspace{1pt}\bbet}\,,
&\
\ERDsig{\mu}{\nu}{\balp}{\dot{\gamma}}
&=
-\,\frac{1}{4}\bigl(\,\EsigD{\mu}{\bet}{\balp}\EDsigD{\nu}{\bet}{\dot{\gamma}}
-\EsigD{\nu}{\bet}{\balp}\EDsigD{\mu}{\bet}{\dot{\gamma}}\,\bigr)\,,
&
\end{align*}
where $\varepsilon_{\alp\hspace{1pt}\gamma}$  $(\varepsilon^{12}=-\varepsilon^{21}=-\varepsilon_{12}=\varepsilon_{21}=1)$ is the spinor metric tensor,
\begin{equation}
\sigma_E=(i\sigma^1, i\sigma^2, i\sigma^3, {1}),
\qquad
\bar{\sigma}_E=(-i\sigma^1,-i\sigma^2,-i\sigma^3,{1})
\nonumber
\end{equation}
and $\sigma^1,\sigma^2,\sigma^3$ are the Pauli matrices. Explicitly
\begin{align}
\Esig{1}{}{}&=\left(\begin{smallmatrix}
                  -i&\phantom{-}0\\
                  \phantom{-}0&\phantom{0}i
                  \end{smallmatrix}\right),
&\                 
\Esig{2}{}{}&=\left(\begin{smallmatrix}
                  -1&\phantom{-}0\\
                  \phantom{-}0&-1
                  \end{smallmatrix}\right),
&\
\Esig{3}{}{}&=\left(\begin{smallmatrix}
                 0& \phantom{0}i\\
                 i&\phantom{0} 0
                  \end{smallmatrix}\right),
&\                 
\Esig{4}{}{}&=\left(\begin{smallmatrix}
                  0&-1\\
                  1&\phantom{-}0
                  \end{smallmatrix}\right),
\nonumber\\[1em]
\EDsig{1}{}{}&=\left(\begin{smallmatrix}
                  i&\phantom{-}0\\
                  0&-i
                  \end{smallmatrix}\right),
&\               
\EDsig{2}{}{}&=\left(\begin{smallmatrix}
                  -1&\phantom{-}0\\
                  \phantom{-}0&-1
                  \end{smallmatrix}\right),
&\
\EDsig{3}{}{}&=\left(\begin{smallmatrix}
                  \phantom{-}0&-i\\
                  -i&\phantom{-}0
                 \end{smallmatrix}\right),
&\
\EDsig{4}{}{}&=\left(\begin{smallmatrix}
                  \phantom{-}0&\phantom{-}1\\
                   -1&\phantom{-}0
                  \end{smallmatrix}\right),
                  \nonumber
\end{align}
\begin{align}
\ERsig{1}{2}{}{}&=\tfrac{1}{2}\left(\begin{smallmatrix}
                  i&\phantom{-}0\\
                  0&-i
                  \end{smallmatrix}\right),
&\                 
\ERsig{1}{3}{}{}&=\tfrac{1}{2}\left(\begin{smallmatrix}
                  \phantom{-}0&\phantom{-}1\\
                 -1&\phantom{-}0
                  \end{smallmatrix}\right),
&\
\ERsig{1}{4}{}{}&=\tfrac{1}{2}\left(\begin{smallmatrix}
                 0& \phantom{0}i\\
                 i&\phantom{0} 0
                  \end{smallmatrix}\right),
\nonumber\\[1em]
\ERsig{2}{3}{}{}&=\tfrac{1}{2}\left(\begin{smallmatrix}
                 0& \phantom{0}i\\
                 i&\phantom{0} 0
                  \end{smallmatrix}\right),
&\                  
\ERsig{2}{4}{}{}&=\tfrac{1}{2}\left(\begin{smallmatrix}
                 0& -1\\
                 1&\phantom{0} 0
                  \end{smallmatrix}\right),
&\
\ERsig{3}{4}{}{}&=\tfrac{1}{2}\left(\begin{smallmatrix}
                  i&\phantom{-}0\\
                  0&-i
                  \end{smallmatrix}\right),
                  \nonumber
\end{align}
\begin{align}
\ERDsig{1}{2}{}{}&=\tfrac{1}{2}\left(\begin{smallmatrix}
                  -i&\phantom{-}0\\
                  \phantom{-}0&\phantom{-}i
                  \end{smallmatrix}\right),
&\                 
\ERDsig{1}{3}{}{}&=\tfrac{1}{2}\left(\begin{smallmatrix}
                  \phantom{-}0&\phantom{-}1\\
                 -1&\phantom{-}0
                  \end{smallmatrix}\right),
&\
\ERDsig{1}{4}{}{}&=\tfrac{1}{2}\left(\begin{smallmatrix}
                 0& \phantom{0}i\\
                 i&\phantom{0} 0
                  \end{smallmatrix}\right),
\nonumber\\[1em]
\ERDsig{2}{3}{}{}&=\tfrac{1}{2}\left(\begin{smallmatrix}
                \phantom{0}0& -i\\
                 -i&\phantom{0} 0
                  \end{smallmatrix}\right),
&\                  
\ERDsig{2}{4}{}{}&=\tfrac{1}{2}\left(\begin{smallmatrix}
                 \phantom{0}0&\phantom{0} 1\\
                 -1&\phantom{0} 0
                  \end{smallmatrix}\right),
&\
\ERDsig{3}{4}{}{}&=\tfrac{1}{2}\left(\begin{smallmatrix}
                  i&\phantom{-}0\\
                  0&-i
                  \end{smallmatrix}\right).\label{sigma_realization}
\end{align}

Relations similar to the ones for the Pauli matrices hold for the $\Sigma$-s \cite[Appendix]{N15}.

\medskip

b) The $N$-extended complex superconformal algebra is isomorphic to $sl(4|N)$ superalgebra, for $N\neq4$.
   Generators: even -- $\E{\alp}{\bet},\,\E{\balp}{\bbet},\,\E{\alp}{\bbet},\,\E{\balp}{\bet}$ (generate $sl(4)$ part), $\E{\Ia}{\Ib}$ (generate $sl(N)$ part), $Y$ ($U(1)$ part);
   odd  --  $\E{\alp}{\Ia}$, $\E{\balp}{\Ia}$, $\E{\Ia}{\alp}$, $\E{\Ia}{\balp}$, $\alp,\bet,\balp,\bbet=1,2$, $\Ia,\Ib=1,2,\ldots,N$.
   
   Let $e^i_j$, $i,j=1,2,\ldots,4+N$ are $(4+N)\times(4+N)$ matrices with entries $(e^i_j)^k_l=\delta^{ik}\delta_{jl}$. Let  $\dot{1}\leftrightarrow 3, \dot{2}\leftrightarrow 4$, when $\balp,\bbet,\dot{\gamma},\ldots$ are used with the $e$-s. Then the matrix realization is
\begin{gather}
\E{\alp}{\bet}=-e^\alp_\bet +\frac{1}{4}\delt{\alp}{\bet}(\sum_{\gamma}e^\gamma_\gamma +\sum_{\dot{\gamma}}e^{\dot{\gamma}}_{\dot{\gamma}})\,,
\quad
\E{\balp}{\bbet}=-e^{\balp}_{\bbet} +\frac{1}{4}\delt{\balp}{\bbet}(\sum_{\gamma}e^\gamma_\gamma +\sum_{\dot{\gamma}}e^{\dot{\gamma}}_{\dot{\gamma}})\,,
\nonumber\\
\E{\alp}{\bbet}=-e^{\alp}_{\bbet}\,,
\quad
\E{\balp}{\bet}=-e^{\balp}_{\bet}\,,
\quad
\E{\Ia}{\Ib}= -e^{\Ia+4}_{\Ib+4} + \frac{1}{N}\delt{\Ia}{\Ib}\sum_{C}e^{C+4}_{C+4}\,,
\nonumber\\
\E{\alp}{\Ia}= -e^{\alp}_{\Ia+4}\,,
\quad
\E{\balp}{\Ia}= -e^{\balp}_{\Ia+4}\,,
\quad
\E{\Ia}{\alp}= -e_{\alp}^{\Ia+4}\,,
\quad
\E{\Ia}{\balp}= -e_{\balp}^{\Ia+4}\,,
\nonumber\\
Y
=
-\frac{1}{4-N} \bigl ( N(\sum_{\gamma}e^\gamma_\gamma +\sum_{\dot{\gamma}}e^{\dot{\gamma}}_{\dot{\gamma}})
+ 
4 \sum_{C}e^{C+4}_{C+4} \bigr)\,.
\label{matrix_realization_of_sl(4|N)_generators}
\end{gather}
Generators $\E{\alp}{\bet},\,\E{\balp}{\bbet},\,\E{\alp}{\bbet},\,\E{\balp}{\bet}$ satisfy commutation relations \eqref{sl_CR} and the relation \eqref{tr_sl_zero}. The other nonzero (graded) commutators are:
\begin{gather*}
\com{\E{\Ia}{\Ib}}{\E{C}{D}}
=
\delt{\Ia}{D}\E{C}{\Ib} 
\,-\,
\delt{C}{\Ib}  \E{\Ia}{D}\,,
\end{gather*}
\begin{align*}
\com{\E{\alp}{\bet}}{\E{\gamma}{\Ia}}
&=
-\,\delta^{\gamma}_{\bet}\,\E{\alp}{\Ia}
+\,\frac{1}{4} \delt{\alp}{\bet} \E{\gamma}{\Ia},
&\qquad
\com{\E{\alp}{\bet}}{\E{\Ia}{\gamma}}
&=
\delta^{\alp}_{\gamma}\,\E{\Ia}{\bet}
-\,\frac{1}{4} \delt{\alp}{\bet} \E{\Ia}{\gamma},
\\[0.45em]
\com{\E{\balp}{\bbet}}{\E{\gamma}{\Ia}}
&=
\frac{1}{4} \delt{\balp}{\bbet} \E{\gamma}{\Ia},
&\qquad
\com{\E{\balp}{\bbet}}{\E{\Ia}{\gamma}}
&=
-\,\frac{1}{4} \delt{\balp}{\bbet} \E{\Ia}{\gamma},
\\[0.45em]
\com{\E{\balp}{\bet}}{\E{\gamma}{\Ia}}
&=
-\,\delta^{\gamma}_{\bet}\,\E{\balp}{\Ia},
&\qquad
\com{\E{\alp}{\bbet}}{\E{\Ia}{\gamma}}
&=
\delt{\alp}{\gamma}\,\E{\Ia}{\bbet},
\\[0.45em]
\com{\E{\alp}{\bet}}{\E{\bgamma}{\Ia}}
&=
\frac{1}{4} \delt{\alp}{\bet} \E{\bgamma}{\Ia},
&\qquad
\com{\E{\alp}{\bet}}{\E{\Ia}{\bgamma}}
&=
-\,\frac{1}{4} \delt{\alp}{\bet} \E{\Ia}{\bgamma},
\\[0.45em]
\com{\E{\balp}{\bbet}}{\E{\bgamma}{\Ia}}
&=
-\,\delt{\bgamma}{\bbet}\,\E{\balp}{\Ia}
+\,\frac{1}{4} \delt{\balp}{\bbet} \E{\bgamma}{\Ia},
&\qquad
\com{\E{\balp}{\bbet}}{\E{\Ia}{\bgamma}}
&=
\delt{\balp}{\bgamma}\,\E{\Ia}{\bbet}
-\,\frac{1}{4} \delt{\balp}{\bbet} \E{\Ia}{\bgamma},
\\[0.45em]
\com{\E{\alp}{\bbet}}{\E{\bgamma}{\Ia}}
&=
-\, \delt{\bgamma}{\bbet}\E{\alp}{\Ia},
&\quad
\com{\E{\balp}{\bet}}{\E{\Ia}{\bgamma}}
&=
\delt{\balp}{\bgamma}\E{\Ia}{\bet}\,,
\\[1.5em]
\com{\E{\Ia}{\Ib}}{\E{\gamma}{C}}
&=
\delt{\Ia}{C}\E{\gamma}{\Ib}
-\,\frac{1}{N}\delt{\Ia}{\Ib}\E{\gamma}{C}\,,
&\qquad
\com{\E{\Ia}{\Ib}}{\E{C}{\gamma}}
&=
-\,\delt{C}{\Ib}\E{\Ia}{\gamma}
+\,\frac{1}{N}\delt{\Ia}{\Ib}\E{C}{\gamma}\,,
\\[0.45em]
\com{\E{\Ia}{\Ib}}{\E{\bgamma}{C}}
&=
\delt{\Ia}{C}\E{\bgamma}{\Ib}
-\,\frac{1}{N}\delt{\Ia}{\Ib}\E{\bgamma}{C}\,,
&\qquad
\com{\E{\Ia}{\Ib}}{\E{C}{\bgamma}}
&=
-\,\delt{C}{\Ib}\E{\Ia}{\bgamma}
+\,\frac{1}{N}\delt{\Ia}{\Ib}\E{C}{\bgamma}\,,
\end{align*}
\begin{align*}
\com{Y}{\E{\gamma}{C}}
=
\E{\gamma}{C}\,,
&\quad
\com{Y}{\E{C}{\gamma}}
=
-\,\E{C}{\gamma}\,,
&\quad
\com{Y}{\E{\bgamma}{C}}
=
\E{\bgamma}{C}\,,
&\quad
\com{Y}{\E{C}{\bgamma}}
=
-\,\E{C}{\bgamma}\,,
\end{align*}
\begin{align}
\com{\E{\Ia}{\alp}}{\E{\gamma}{C}}
&=
-\,\delt{\Ia}{C}\E{\gamma}{\alp}
-\,
\delt{\gamma} {\alp}\E{\Ia}{C}
-\,
\frac{4-N}{4N}\delt{\Ia}{C}\delt{\gamma}{\alp} Y\,,
&\quad
\com{\E{\balp}{\Ia}}{\E{C}{\gamma}}
&=
-\,\delt{C}{\Ia}\E{\balp}{\gamma}\,,
&\nonumber\\[0.45em]
\com{\E{\Ia}{\balp}}{\E{\bgamma}{C}}
&=
-\,\delt{\Ia}{C}\E{\bgamma}{\balp}
-\,
\delt{\bgamma} {\balp}\E{\Ia}{C}
-\,
\frac{4-N}{4N}\delt{\Ia}{C}\delt{\bgamma}{\balp} Y\,,
&\quad
\com{\E{\alp}{\Ia}}{\E{C}{\bgamma}}
&=
-\,\delt{C}{\Ia}\E{\alp}{\bgamma}\,.
\end{align}
The relation with the generators from a) is:
\begin{gather}
\EROT{\mu}{\nu}
\,=\,
\ERsig{\mu}{\nu}{\alp}{\bet}\,\E{\bet}{\alp}
\,+\,
\ERDsig{\mu}{\nu}{\balp}{\bbet}\,\E{\bbet}{\balp}\,,
\qquad
\EDL
\,=\,
\frac{1}{2}\,\bigl(\,\sum_{\alp=1}^{2}\,\E{\alp}{\alp}
\,-\,
\sum_{\balp=1}^{2}\,\E{\balp}{\balp}\,\bigr)\,,
\nonumber\\
\ETR_{\mu}
\,=\,
-
\,\EDsigD{\mu}{\alp}{\bbet}\,\E{\bbet}{\alp}\,,
\qquad
\ESC_{\mu}
\,=\,
-\,
\EsigD{\mu}{\bet}{\balp}\,\E{\bet}{\balp}\,,
\qquad
\RSYM^{\Ia}_{\Ib}
\,=\,
\E{\Ia}{\Ib}\,,
\qquad
\RCHA
\,=\,
-\frac{1}{2} Y,
\nonumber\\[0.5em]
\EST{\alp}{\Ia}
\,=\,
2\E{\Ia}{\alp}\,,
\qquad
\ESSC{\alp}{\Ia}
\,=\,
2\E{\alp}{\Ia},
\qquad
\EBST{\balp}{\Ia}
\,=\,
\,2\E{\balp}{\Ia}\,,
\qquad
\EBSS{\balp}{\Ia}
\,=\,
2 \E{\Ia}{\balp}\,.
\label{superconformal_to_sl(4)}
\end{gather}

\medskip
Remark: For $N=4$, the $sl(4|4)$ superalgebra is not simple, it contains an one-dimensional ideal $I$, generated by identity matrix (which is an  $sl(4|4)$-matrix).    In this case the extended superconformal algebra is isomorphic to $psl(4|4):=sl(4|4)/I$.

\end{document}